\documentclass[11pt]{article}
\usepackage[utf8]{inputenc}
\usepackage{amsmath}
\usepackage{amssymb}
\usepackage[margin=0.75in]{geometry}
\usepackage{hyperref}
\usepackage{cleveref}
\usepackage{indentfirst}
\usepackage{titlesec}
\usepackage{graphicx}
\usepackage[dvipsnames]{xcolor}

\newcommand{\del}{\partial}

\newcommand{\m}{\mu}
\newcommand{\n}{\nu}

\newcommand{\vp}{\varphi}

\newcommand{\td}{\text{d}}

\newcommand{\isEquivTo}[1]{\underset{#1}{\sim}}

\title {\bf Disforming the Kerr metric}
\date{}
\author{Timothy Anson $^{1}$, Eugeny Babichev $^{1}$, Christos Charmousis $^{1}$ and Mokhtar Hassaine $^{2}$\\
    $^{1}$ Universit\'e Paris-Saclay, CNRS/IN2P3, IJCLab, 91405 Orsay, France \\
    $^{2}$ Instituto de Matem\'{a}tica y F\'{\i}sica, Universidad de
    Talca, Casilla 747,Talca, Chile.}

\begin{document}

{\let\newpage\relax\maketitle}

\begin{abstract}
Starting from a recently constructed stealth Kerr solution of  higher order scalar tensor theory involving  scalar hair, we analytically construct disformal versions of the Kerr spacetime with a constant degree of disformality and a regular scalar field. While the disformed metric has only a ring singularity and asymptotically is quite similar to Kerr, it is found to be neither Ricci flat nor circular. Non-circularity has far reaching consequences on the structure of the solution. As we approach the rotating compact object from asymptotic infinity we find a static limit ergosurface similar to the Kerr spacetime with an enclosed ergoregion. However, the stationary limit of infalling observers is found to be a timelike hypersurface. A candidate event horizon is found in the interior of this stationary limit surface. It is a null hypersurface generated by a null congruence of light rays which are no longer Killing vectors. Under a mild regularity assumption, we find that the candidate surface is indeed an event horizon and the disformed Kerr metric is therefore a black hole quite distinct from the Kerr solution.
\end{abstract}

{\it{We humbly dedicate this work to the memory of our colleague Renaud Parentani. }}

\section{Introduction}

There is a plethora of novel compelling evidence, from gravitational wave emission of distant binaries \cite{LIGOScientific:2018mvr,Abbott:2016blz},  to the Event Horizon Telescope \cite{Johannsen:2016vqy,Doeleman:2009te} and the instrument GRAVITY \cite{ Abuter:2018drb} for supermassive compact objects, that black holes exist and are rotating. In General Relativity (GR) a stationary and axially symmetric rotating black hole is described by the Kerr metric \cite{Kerr:1963ud}. The Kerr metric is the unique vacuum black hole with the afore mentioned symmetries in GR \cite{Robinson:1975bv}. Therefore it is fair to say that the Kerr solution is probably the most important of solutions to the Einstein equations. Although a rather complex metric, given that it is a solution of partial differential equations, it has a number of hidden symmetries and mathematical properties that make physical applications tractable (and even analytically so to a certain extent).

For a start, it is already quite a feat that one can analytically solve the partial differential equations in this case \cite{Kerr:1963ud}, as well as in the case of a cosmological constant \cite{CARTER1968399}. In fact it is only in 4 dimensions that we can analytically find the relevant charged solution \cite{Newman:1965my}, whereas the plethora of higher dimensional rotating solutions have only recently been worked out in full \cite{Gibbons:2004uw}. Furthermore, the Kerr geodesics can be computed analytically because the relevant Hamilton-Jacobi function is found to be separable \cite{Carter:1968ks}, a fact that is associated to the existence of an additional Killing tensor \cite{Walker:1970un}. Making use of the Newman-Penrose formalism, linear perturbations of the Kerr spacetime can be written in a separable form, the Teukolsky equation \cite{Teukolsky:1973ha}. These mathematical properties pave the way to understanding, amongst other things, linear stability and quasi-normal modes (which are important for the ring down phase of binaries) and then geodesics for black hole shadows.
The presence of an ergosphere region, where static observers cease to exist, results in fascinating effects such as superradiance~\cite{superradiance} and the Penrose process of extracting energy from black hole rotation~\cite{Penrose:1969}.
These are part of a wide class of physical phenomena, which also include the quantum laser effects of black holes \cite{Corley:1998rk,Leonhardt:2008js,Coutant:2009cu} (and their acoustic counterparts \cite{Finazzi:2010nc}). Hence it is fair to say that the Kerr solution is not only phenomenologically important but also of major theoretical interest.

All observational data up to now agrees with predictions of the Kerr metric. In light of recent/future observational advances, it becomes crucial to find competing backgrounds to the GR prototype solution.
However, it is rather difficult to produce solutions, analytically or even numerically, which have a similar geometry while remaining distinctively different (see e.g. \cite{Papadopoulos:2018nvd,Johannsen:2013rqa,Johannsen:2015pca} for some interesting geometrical approaches to construct deviations from the Kerr metric, mainly of ``phenomenological'' origin, i.e. those metrics are not solutions of any particular gravity theory but rather maintain certain properties of the Kerr spacetime). Interesting hairy numerical constructions have been obtained by considering scalar or other forms of matter (including for example \cite{Herdeiro:2014goa,Grandclement:2014msa,Volkov:1998cc,Kleihaus:2000kg}). The object of this article is to construct a specific analytic solution of modified gravity with the same spacetime symmetries as Kerr (stationary and axisymmetric), which is sufficiently similar yet distinctively different. Our starting point will be degenerate (DHOST or EST) scalar tensor theories \cite{Langlois:2015cwa,Crisostomi:2016czh,BenAchour:2016fzp}, which are the working prototype of modified gravity theories with a single additional degree of freedom. Although the solution we present is an exact solution of a precise theory belonging to this class, we believe that the solution can be of more interest than the theory itself as it gives a precise analytic form of a different, yet competing metric to the prototype GR solution. The solution we shall present also singles out a number of properties that simply cease to exist once we venture away from GR. Some of these properties or shortcomings can be of theoretical interest for black hole physics, for example horizon properties, thermodynamics, etc. Even from the point of view of GR, it is enlightening to discuss metrics which are counterexamples to usual black hole properties. This is somewhat similar to the Taub-NUT spacetime in GR being ``a counterexample to almost anything \cite{Misner:1965zz}''.

Recently, using the geodesic paradigm of Carter's seminal work \cite{Carter:1968ks}, it was understood that one could ``paint'' the Kerr spacetime with well defined scalar hair in a regular fashion \cite{Charmousis:2019vnf}. The theory hosting this solution is a particular DHOST/EST theory, whereas the scalar field is a particular Hamilton-Jacobi function which, crucially, is regular everywhere (including the event horizon). The kinetic term of the scalar is therefore constant, in adequacy with the standard Hamilton Jacobi method for obtaining geodesics (see for example \cite{Misner:1974qy} page 897). Unsurprisingly, this scalar tensor solution has similar properties to the GR Kerr metric. For example, the relevant modified Teukolsky operator for tensor perturbations is again shown to be separable with non observable differences to its GR version \cite{Charmousis:2019fre}. The solution may also present pathologies in the scalar perturbations \cite{Babichev:2018uiw,deRham:2019gha} (one expects to make certain starting concessions in order to obtain analytical solutions for modified gravity). The key to go further are disformal transformations, and in particular those originating from ``geodesic'' scalars.

It is known that disformal and conformal transformations of scalar tensor theories \cite{Zumalacarregui:2013pma,Bettoni:2013diz} are internal maps of the theory, in other words such transformations will take us from some DHOST Ia theory to some other specific DHOST Ia theory (see for example \cite{Achour:2016rkg}). The interesting combination which we believe crucial here is that the scalar responsible for the disformed metric  is related to the geodesics of Kerr. As we will see, the transformation is found to be surprisingly regular yet non trivial. In fact it has been shown that in the static case, for $X$ constant, the disformal version of a Schwarzschild type solution is again a mass rescaled Schwarzschild black hole. With a basic disentangling of coordinates, and given that the disformal factor $B=B(X)$ is a function of $X$ (which is constant on-shell), the resulting metric is geometrically identical with a rescaled mass (and cosmological constant) \cite{Babichev:2017lmw, BenAchour:2019fdf} (see also \cite{Domenech:2019syf} for interesting extensions and questions concerning the regularity of solutions). As we will see in the present paper, once  rotation is present the static picture is completely changed. The disformed Kerr metric is no longer Ricci flat. It is furthermore not even a circular spacetime, as all Einstein metrics are in GR. In a nutshell this means that the metric is no longer reflection symmetric in the $t$ and $\varphi$ Killing coordinates as is the Kerr metric{\footnote{This seemingly bland statement signifies that a rotating black hole rotating in one direction is not necessarily rotating in the opposite direction in its past \cite{Hawking:1973uf}.}}.However, it again has a single ring singularity and is asymptotically very similar to Kerr. It also has an ergosurface beyond which static observers do not exist. Interestingly, the boundary of constant $r$ and $\theta$ stationary observers, which is a Killing/event horizon for Kerr, fails to be so for the disformed metric. The latter hypersurface is not null but actually timelike. Therefore, unlike for the Kerr spacetime, there is an additional stationary limit surface in the ergoregion, inside which the  Killing vectors are spacelike. In fact, there exist special timelike observers (for which $r$ and $\theta$ are not constant) that can venture further in up to the (candidate) event horizon. The generator of this null surface of no return, our (candidate) event horizon, is no longer  a Killing vector. The usual stationary Killing vector of Kerr is now a spacetime pointing vector (apart from the poles). Indeed, these 3 successive hypersurfaces meet at the north and south poles of the rotating solution, so that it is regular there.

The paper is organised as follows: in the next section we will proceed to construct the disformed Kerr solution. While doing so we will remind the reader of some well known properties of the Kerr solution. We will discuss some general properties of the disformed metric in section 3, while in section 4 we will study the special hypersurfaces of the metric:  the ergosurface, the stationary limit and horizons. We will discuss our findings and conclude in the 5th section.

\section{Constructing the disformal transformation of the Kerr metric}
\label{sec:disformal}
We start by constructing an explicit example of a
{\it disformal Kerr metric}. By disformal Kerr metric, we mean a
spacetime metric $\tilde{g}^{\mbox{\tiny{disf}}}_{\mu\nu}$ which can
be represented in the following way:
\begin{eqnarray}
\tilde{g}^{\mbox{\tiny{disf}}}_{\mu\nu}={g}^{\mbox{\tiny{Kerr}}}_{\mu\nu}+B
(\phi,\partial_{\sigma}\phi\partial^{\sigma}\phi)\phi_{\mu}\phi_{\nu}\; ,
\label{disf}
\end{eqnarray}
where $({g}^{\mbox{\tiny{Kerr}}}_{\mu\nu}, \phi)$ is a nontrivial
solution of a subclass of DHOST theory whose metric solution
${g}^{\mbox{\tiny{Kerr}}}_{\mu\nu}$ is the Kerr
spacetime. Our starting block will be the stealth black hole solution found in
\cite{Charmousis:2019vnf} where the authors consider a subclass of DHOST Ia theory, and show that a nontrivial
scalar field defined on the Kerr metric solves the equations of
motion. The scalar field is defined as some particular Hamilton Jacobi function giving a regular geodesic congruence for the Kerr spacetime. It is precisely this geometrically induced scalar field that we will consider to
construct the disformal Kerr metric (\ref{disf})\footnote{Note that stealth black holes were found to have a problem with perturbations, namely the scalar mode becomes non-dynamical on such background solutions, which is related to strong coupling~\cite{Babichev:2018uiw,deRham:2019gha}.
However, it has been argued in~\cite{Motohashi:2019ymr} that  a small detuning of the degeneracy condition for the theory can be used to solve this problem. 
 }. 

In order to be as
self-contained as possible, let us now detail this construction. We
start from the Kerr metric written in Boyer-Lindquist
coordinates,
\begin{equation}
\label{BL} g_{\mu\nu}\td x^\mu \td x^\nu =
-\left(1-\frac{2Mr}{\rho^2}\right)\td t^2
-\frac{4Mar\sin^2\theta}{\rho^2}\td t\td\vp+
\frac{\sin^2\theta}{\rho^2}\left[\left(r^2+a^2\right)^2-a^2\Delta
\sin^2\theta\right]\td\vp^2 + \frac{\rho^2}{\Delta}\td r^2 + \rho^2
\td \theta^2\; ,
\end{equation}
where $M$ represents the mass of the black hole, $a$ is the angular
momentum per unit mass, and for clarity we have defined
\begin{align*}
\Delta &= r^2+a^2 - 2 Mr\; ,\\
\rho^2 &= r^2 + a^2 \cos^2\theta\; .
\end{align*}
Working with Boyer-Lindquist coordinates is mainly
motivated by the minimal number of off-diagonal components, considerably facilitating the calculations. Furthermore, the Killing vectors are adapted to the coordinates and read $\partial_t$ and $\partial_\varphi$.
The event horizons of the original metric are located at constant values of $r$, and are obtained by solving the equation $\Delta=0$,
\begin{equation}
\label{rpm} r_\pm = M\pm \sqrt{M^2-a^2}\; ,
\end{equation}
where $r_+$ is the outer (event) horizon and $r_-$ is the inner
horizon. Extremality occurs for $a=M$, where the two horizons coincide. The black hole event horizon is a Killing horizon, and the stationary observers with $r$ and $\theta$ constant become null there. Another hypersurface of interest is the ergosurface, whose
locus is determined by the equation $g_{tt}=0$, and which is the endpoint of static observers ($r,\theta,\varphi$ constant). This hypersurface is defined by
\begin{equation}
r_E = M + \sqrt{M^2-a^2 \cos^2\theta}\; ,
\end{equation}
and is accessible to far away observers. Note that the two hypersurfaces $r=r_E$ and $r=r_+$ coincide at the poles. The region of spacetime in between the ergosurface and the outer event horizon is the ergoregion of the original Kerr black hole, where one can have interesting physical effects such as the Penrose process. In this region the gravitational pull is so strong that a stationary observer can only rotate along with the black hole (otherwise his future pointing tangent vector could not be timelike). We will see later the rather interesting influence of the disformal transformation on the location of these two hypersurfaces.
There is a single curvature singularity situated at $\rho=0$, while closed timelike curves can develop within this region (where the axial Killing vector becomes timelike $g_{\varphi\varphi}<0$).
There are of course many other important properties (physical or mathematical)
underlying the Kerr solution, and it proves interesting
to study the impact of the disformal transformation on them. We will encounter some of them in the forthcoming sections.

In~\cite{Charmousis:2019vnf}, it was shown that the scalar hair painting (\ref{BL}) is given by
\begin{equation}
\label{phi} \phi = q \left[t + \int \frac{\sqrt{2M
r(a^2+r^2)}}{\Delta}\td r\right]\;,
\end{equation}
where $q$ is constant that we assume positive. Note that the relative $+$ sign  was
chosen for the scalar field to be regular at the Kerr horizons
$\Delta=0$. Apart from being linear in time, the scalar field has a
constant standard kinetic term:
\begin{equation}
\label{Xcst} 
X\equiv g^{\mu\nu}\partial_\mu\phi\partial_\nu\phi=-q^2\; .
\end{equation}
This last equation is nothing but the Hamilton-Jacobi (HJ) equation determining the most general HJ function for Kerr spacetime.
Carter demonstrated that  the geodesic equations are integrable for the Kerr spacetime \cite{Carter:1968ks}, meaning that there is an equal number of conserved quantities and spacetime directions. These quantities are energy at infinity $E$, angular momentum $L$, rest mass $m$ and Carter's separation constant $C$ (whose proof of existence gives integrability). The former two originate from the Killing vectors $\partial_t$ and $\partial_\varphi$, while the latter two come from the existence of Killing tensors for the Kerr spacetime.
In order for our scalar hair to be well defined from the event horizon up to asymptotic infinity, one must take $E=m=q$ (where particles can marginally reach timelike infinity), and $L=C=0$ (to have regularity at the poles).

We now have all the necessary ingredients to perform the disformal
transformation as defined by Eq. (\ref{disf}). Since the DHOST
theory admitting  the stealth Kerr solution is invariant
under the transformation $\phi\to\phi+\mbox{cst}$, it is natural to
consider a disformal function $B$ with the same symmetry, i.
e.  $B=B(X)$. On the other hand,  $X$ being constant (\ref{Xcst}),
the disformal Kerr metric associated to the Kerr stealth solution
\cite{Charmousis:2019vnf} can be written without any loss of
generality as
\begin{equation}
\label{disformal} \tilde{g}_{\m\n} = g_{\m\n} - \frac{D}{q^2}\;
\del_\mu\phi\,\del_\nu\phi\; ,
\end{equation}
where $D$ is a constant whose sign is not fixed a priori.
In the original coordinates, the disformal Kerr metric reads
\begin{equation}
\begin{split}
\label{df}
\tilde{g}_{\mu\nu}\td x^\mu \td x^\nu &= -\left(1-\frac{2\tilde{M}r}{\rho^2}\right)\td t^2 -\frac{4\sqrt{1+D}\tilde{M}ar\sin^2\theta}{\rho^2}\td t\td\vp + \frac{\sin^2\theta}{\rho^2}\left[\left(r^2+a^2\right)^2-a^2\Delta
\sin^2\theta\right]\td\vp^2\\
 &+  \frac{\rho^2 \Delta - 2 \tilde{M}(1+D) r D (a^2+r^2)}{\Delta^2}\td r^2 - 2D \frac{\sqrt{2\tilde{M}r(a^2+r^2)}}{\Delta}\td t\td r + \rho^2 \td \theta^2\; .
\end{split}
\end{equation}
For simplicity, we have introduced the rescaled mass $\tilde{M} =
M/(1+D)$, and the coordinate $t$ has been
conveniently rescaled as $t\to t/ \sqrt{1+D}$, assuming $1+D>0$. In these coordinates, the determinant of the disformal metric is the same as that of the Kerr metric:
\begin{equation}
\nonumber \sqrt{-\tilde{g}}=\sqrt{-g}\; .
\end{equation}
We will be referring to $D$ as the metric's disformality. Indeed, although $D$ is a given constant parametrising a given (DHOST Ia) theory, in a more phenomenological approach, it can be thought of as an additional parameter of the disformed metric. Given that $\tilde{X} =-q^2/(1+D)$ the scalar field is again a HJ function for the disformed metric.

Before entering the details of the properties of the disformal
metric, we would like to emphasize that its nontrivial character is
mainly due to the time dependence of the scalar field conjugated
with the non-zero angular momentum $a$.
A non-trivial property of the new metric is the term $\tilde{g}_{tr}$, which cannot be eliminated by a coordinate change without introducing other off-diagonal elements (see discussion in Sec.~\ref{sec:properties}).
This is one of the main differences with the Kerr metric, and we will see below that the presence of this extra
off-diagonal term will have
significant consequences. On the other hand, in the static limit
case $a=0$, the off-diagonal term $\tilde{g}_{tr}$
in~(\ref{df}) can be removed by the following coordinate
transformation:
\begin{equation}
\td t = \td T -
\frac{D\sqrt{2\tilde{M}r^3}}{\Delta\left(1-\frac{2\tilde{M}}{r}\right)}\td
r\; ,
\end{equation}
and the resulting metric is nothing but the Schwarzschild spacetime
with a rescaled mass:
\begin{equation*}
\tilde{g}_{\mu\nu}\td x^\mu \td x^\nu =
-\left(1-\frac{2\tilde{M}}{r}\right)\td T^2 +
\left(1-\frac{2\tilde{M}}{r}\right)^{-1}\td r^2 + r^2\td \Omega^2\;
.
\end{equation*}
Hence, in the Schwarzschild  static case, the net effect of the
disformal transformation is only to rescale the mass parameter. Note
that a similar observation has been noted in the case of
Schwarzschild-de-Sitter metric in
Ref.~\cite{Babichev:2017lmw,Babichev:2018uiw, BenAchour:2019fdf} (see also \cite{Dubovsky:2006vk}).

\section{General properties of the disformed  Kerr metric}
\label{sec:properties}
Let us analyze some of the properties
of the disformal Kerr metric (\ref{df}), starting from the
singularities.
An initial inspection of the metric may naively lead to the conclusion that the
hypersurface defined by $r=r_\pm$ is singular.
Nevertheless, as in GR, this latter hypersurface is merely a coordinate
singularity, which can be removed by an appropriate choice of
coordinates. One can start by computing metric
scalar invariants such as
\begin{equation*}
\tilde{R} = -\frac{Da^2 Mr [1 + 3
\cos(2\theta)]}{(1+D)\rho^6},\;\;\;
\tilde{R}_{\mu\nu}\tilde{R}^{\mu\nu}= \frac{D^2a^4
M^2Q_1(r,\theta)}{4\rho^{12}(r^2+a^2)(1+D)^2},\;\;\;
\tilde{R}_{\mu\nu\alpha\beta}\tilde{R}^{\mu\nu\alpha\beta} = \frac{
M^2Q_2(r,\theta)}{\rho^{12}(r^2+a^2)(1+D)^2}\; ,
\end{equation*}
where  the expressions of $Q_1$ and $Q_2$ can be found in the
Appendix~\ref{AppendixQ}. Although the regularity of scalar invariants does not constitute a necessary and sufficient condition for regularity of the metric it is nevertheless a good starting point.
The above expressions suggest that there are no physical
singularities apart from the standard ring singularity at $\rho=0$.
A more rigorous argument is given in the
Appendix~\ref{AppendixKerr} where the disformal spacetime (\ref{df})
is re-written in a Kerr coordinate system making apparent its regularity
at $r=r_\pm$.
In this coordinate system, both the metric and the scalar field are manifestly
well-defined except at the ring singularity $\rho=0$.
We will deal with possible horizon hypersurfaces and coordinate singularites in the next section.

It is clear that the disformal Kerr metric (\ref{df}) again represents a stationary
and axially symmetric spacetime, whose Killing vector fields $\partial_t$
and $\partial_{\vp}$ are manifest since the metric coefficients are
independent of $t$ and $\vp$. In addition to these Killing vectors, the Kerr metric possesses a non-trivial Killing tensor $K_{\mu\nu}$ which verifies the Killing equation $\nabla_{(\alpha} K_{\mu\nu)}=0$. This gives rise to the conservation of Carter's constant along geodesics, and enables the integration of the geodesic equation. For small deformations $D\ll1$, one can try to construct a symmetric tensor of the form $\tilde{K}_{\mu\nu} = K_{\mu\nu} + D \delta K_{\mu\nu}$ which satisfies the linearized Killing equation (to order 1 in $D$). It can be shown that the system of equations is not consistent. This means that if a Killing tensor does exist for the disformal metric, it cannot be written as a deformation of the Killing tensor $K_{\mu\nu}$ of the Kerr metric.

It is well known that Einstein metrics belonging to the class
of stationary and axisymmetric spacetimes are
circular spacetimes, see e.g.~\cite{Frolov:1998wf}. This means that locally the metric is not only independent
of the time and of the rotation angle but also invariant under the
"$t-\vp$" reflection isometry, which is the simultaneous change of
the time and of the rotation angle. A rotating solution such as a rotating neutron star with toroidal magentic field, for example, may fail to be a circular spacetime (see for example \cite{noncircular}) but it is fair to say that most known solutions in the literature are indeed circular. Rigorously, an axisymmetric
spacetime is said to be circular if the $2-$surfaces orthogonal to
the Killing fields $k$ and $\eta$, respectively the one-forms
associated to the Killing vectors $\del_t$ and $\del_\vp$,
\begin{align*}
k = \tilde{g}_{t\nu}\td x^\nu\; ,\quad  \eta = \tilde{g}_{\vp\nu}\td x^\nu\; ,
\end{align*}
are integrable. According to Frobenius's theorem, the circularity of
the metric is equivalent to the following integrability conditions:
\begin{equation*}
k\wedge\eta\wedge\td k = k\wedge\eta\wedge\td\eta=0 \; .
\end{equation*}
In our case, an explicit calculation yields
\begin{eqnarray}
k\wedge\eta\wedge\td k &=&
\tilde{g}_{tr}\left(\tilde{g}_{\vp\vp}\frac{\del
\tilde{g}_{tt}}{\del\theta}-
\tilde{g}_{t\vp}\frac{\del \tilde{g}_{t\vp}}{\del\theta}\right) \td t\wedge\td r\wedge \td \theta\wedge\td\vp\nonumber\\
&=& -D\frac{4 a^2 \tilde{ M} r \sqrt{2\tilde{M}r(a^2+r^2)}
\cos\theta\sin^3\theta}{\rho^4}\td t\wedge\td r\wedge \td
\theta\wedge\td\vp\; ,\label{froe}
\end{eqnarray}
 and hence the disformal Kerr metric is not circular. In addition, since the
integrability conditions are equivalent to the eliminability of the
cross-terms  $\tilde{g}_{tr}, \tilde{g}_{t\theta},
\tilde{g}_{\vp r}$ and $ \tilde{g}_{\vp\theta}$ (for our choice of
adapted coordinates to the Killing vectors $\partial_t$ and $\partial_\varphi$), our disformal metric can not be cast in the
Lewis-Papapetrou form as in vacuum GR (or with a circular energy-momentum tensor \cite{Papapetrou:1966zz}).
In particular, this implies that the off-diagonal term $\tilde{g}_{tr}$ cannot be removed by a coordinate change without introducing other off-diagonal elements that break circularity. 

The circularity property is twofold: it is known to be fundamental for proving
important theorems such as the constancy of surface gravity, but can also prove to be too
restrictive. An illustrative example of this latter fact are black holes in a theory with a conformally coupled scalar where a non trivial static and spherically symmetric solution exists \cite{bbmb}. This is the familiar BBMB black hole with secondary scalar hair.
It is however proven, under the crucial hypothesis of circularity, that stationary-axisymmetric asymptotically flat black holes in this theory are only those described by the Kerr metric and  trivial scalar hair~\cite{Zannias:1998jz}.
In our case, the lack of
circularity will turn out  to be fundamental. For a start, it is easy to see that the disformal Kerr metric (\ref{df}) is in general not Ricci flat. If it were, circularity would of course follow.
A simple argument comes from the relation of the original and disformal Ricci tensors (\ref{disf}), which reads
\begin{equation}
\label{newRicci} \tilde{R}_{\m\n} - R_{\m\n} =
-\frac{D}{q^2(1+D)}\left[\square\phi\phi_{\mu\nu} -
{\phi^\alpha}_\mu \phi_{\alpha\n} -
R_{\m\alpha\n\beta}\phi^\alpha\phi^\beta\right]\; .
\end{equation}
Hence, since the Kerr metric is Ricci flat,
$R_{\m\n}=0$, the Ricci tensor of the disformal metric is given by
the right hand side of (\ref{newRicci}), which does not in general vanish (it can do only for certain scalars sourcing highly symmetric cases such as spherical symmetry \cite{Babichev:2017lmw,BenAchour:2019fdf}).
Once we have rotation the disformal metric is no longer Ricci flat and  loses the circularity property as well. This inevitably leads to fundamental differences and to the loss of usual basic properties of GR black holes, as we will now see.

Let us study the asymptotic region at large
distance  $r\gg M$ in order to compare the disformal metric
with the Kerr spacetime in this region.  This comparison may be important for
astrophysical applications, in particular by observing orbiting
stars around the supermassive black hole in the center of our
galaxy.
Although in general the off-diagonal term $\tilde{g}_{tr}$ cannot be eliminated, it is possible to do so in the asymptotic region $r\gg M$. Indeed, through a  redefinition of the time coordinate
\begin{equation}
\label{coord_change}
\td t = \td T  -D \frac{\sqrt{2\tilde{ M}r(a^2+r^2)}}{\Delta(1-
\frac{2\tilde{ M}}{r})} \td r\; ,
\end{equation}
the cross term becomes
\begin{equation*}
\tilde{g}_{Tr}=\frac{2 D \tilde{M} a^2 \cos^2\theta\sqrt{2\tilde{M}r(a^2+r^2)}}{(r-2\tilde{M})\rho^2\Delta}\isEquivTo{r\to\infty}\mathcal{O}\,\left(\frac{1}{r^{7/2}}\right)\; ,
\end{equation*}
and one can see that it decays rapidly enough at large $r$. It is then convenient to
rewrite the line element in terms of Cartesian  coordinates $x =
r\sin\theta\cos\vp$, $y=~ r\sin\theta\sin\vp$ and $z = r\cos\theta$.
Keeping only the leading order corrections, the line element of the
disformal Kerr metric in the asymptotic region reads
\begin{equation}
\label{asymptotic} \td \tilde{s}^2  = \td s_\text{Kerr}^2+
\frac{D}{1+D}\left[\mathcal{O}\left(\frac{\tilde{a}^2\tilde{M}}{r^3}\right)\td
T^2 +
\mathcal{O}\left(\frac{\tilde{a}^2\tilde{M}^{3/2}}{r^{7/2}}\right)\alpha_i\td
T\td x^i +
\mathcal{O}\left(\frac{\tilde{a}^2}{r^{2}}\right)\beta_{ij}\td
x^i\td x^j\right]\; .
\end{equation}
In the above expression we have defined
$\tilde{a}=\sqrt{1+D} a$ and the $\alpha_i$'s and $\beta_{ij}$'s are $\mathcal{O}(1)$ coefficients.
The first term in~(\ref{asymptotic})  is the line element for the
Kerr metric with parameters $\tilde{a}$ and $\tilde{M}$. For large $r$, it can be written in the form~\cite{Misner:1974qy}:
\begin{equation}
\nonumber
\begin{split}
\td s^2_\text{Kerr} & = -\left[1-\frac{2\tilde{M}}{r} + \mathcal{O}\left(\frac{1}{r^3}\right)\right]\td T^2 -
\left[\frac{4 \tilde{a} \tilde{M}}{r^3}+ \mathcal{O}\left(\frac{1}{r^5}\right)\right]\left[x \td y - y\td x\right]\td T \\
&+ \left[1 + \mathcal{O}\left(\frac{1}{r}\right) \right]\left[\td
x^2 + \td y^2 + \td z^2\right]\; .
\end{split}
\end{equation}
Notice that the Kerr part of the metric contains the rescaled mass $\tilde{M}$
and rescaled angular parameter $\tilde{a}$, rather than $M$ and
$a$.  As can be seen from the above expressions, the effect of
the disformal transformation at leading order is merely a rescaling of
the mass and the parameter $a$, while for the next-to-leading order
corrections the disformal off diagonal terms $\td\tilde{T}\td x^i$ are larger,
$\mathcal{O}\left(1/r^{7/2}\right)$, than those of the Kerr metric,
$\mathcal{O}\left(1/r^4\right)$.

Finally, we finish this section with an important property that stems from the construction of the metric~(\ref{df}) itself.
It is known that in GR there exist solutions that contain time machines, an example being the Kerr metric with $a>M$~\cite{Carter:1968rr}.
It is important therefore  to show that our disformed Kerr metric does not contain such a pathology. This is achieved by showing that the spacetime with the metric $\tilde{g}_{\mu\nu}$ is stably causal, in other words it remains causal under a small perturbation of the light cone. For a spacetime to be stably causal, it is sufficient to show that there exists a function whose gradient is a future directed timelike vector field~\cite{Wald}. This function can be thought of as a global time. 
In our case there is such a function by construction, the scalar field $\phi(t,r)$ itself. Note that due to~(\ref{Xcst}) it is indeed timelike\footnote{The same argument was used in~\cite{Babichev:2007dw} to show that a black hole with an accreting k-essence field~\cite{Babichev:2006vx} has no closed timelike curves. Similarly to our case, the k-essence field was identified as a global time function in~\cite{Babichev:2007dw}.}.
According to~(\ref{phiKerr}), the scalar is regular for $r>0$. Therefore our spacetime is globally causal, provided that 
the region of the spacetime for some positive $r$ (in particular outside the event horizon) is causally disconnected from the region  $r<0$ (where closed timelike curves are present, similarly to the Kerr case).

\section{From the static to the stationary limit and all the way up to the event horizon}
\label{sec:surfaces}

We are now ready to discuss the properties of important hypersurfaces in the metric~(\ref{df}), which include the timelike static and stationary limiting surfaces as well as the null hypersurface(s), which if present are candidate event horizon(s).
We will move gradually from large $r$ to smaller radii studying interesting surfaces on the way, until we meet a perspective horizon or hit the ring singularity.
In order to make it more intuitive, we make our analysis parallel to that of the Kerr metric, which has been summarized in Section 2.

\subsection{The endpoint of static and stationary observers}
The outermost interesting hypersurface in the case of the Kerr metric is the ergosurface, the limiting hypersurface where the timelike Killing vector $l^\mu_{(t)} = \left(1,0,0,0\right)$ becomes null. This surface is often called the static limit, since static timelike observers (with constant $r, \theta$ and $\vp$), can no longer exist in its interior.
Since the disformed metric~(\ref{df}) has the same Killing vector $l^\mu_{(t)}$ as the Kerr metric, we can define the ergosurface in a similar way, and it corresponds to the surface where $l^\mu_{(t)}$ becomes null:
\begin{equation}
\tilde{g}_{\mu\nu}l^\mu_{(t)} l^\nu_{(t)} =0 \;\;\; \Leftrightarrow \;\;\; \tilde{g}_{tt} = 0.
\end{equation}
From the above we have
\begin{equation}
\label{ergosphere}
r(\theta)^2 + a^2 \cos^2\theta = 2\tilde{M}r(\theta)\; ,
\end{equation}
so that the locus of the surface related to the Killing vector  $l^\mu_{(t)}$ is the same as for the Kerr spacetime with a rescaled mass.
Note, however, that the non-rescaled Kerr parameter $a$ enters Eq.~(\ref{ergosphere}).
Therefore, if the disformed and Kerr metrics are matched at leading order for large $r$ (see the discussion in Sec.~\ref{sec:properties}),
the locus of the ergosphere for the two does not coincide. Indeed, the equation for the ergosphere for the Kerr metric (matching the disformed metric at large radii) is of the form (\ref{ergosphere}) with $a\to \tilde{a}$.
We can thus see that the locus of the ergosphere, corresponding to the Killing vector $\partial_t$, is modified with respect to the Kerr case. We refer to the surface~(\ref{ergosphere}) as ergosurface, similarly to the Kerr case, or as static limit.

The next step is to consider a combination of the two independent Killing vectors $\partial_t$ and $\partial_\varphi$:
\begin{equation}
\label{stationary}
l=\del_t + \omega\del_\vp\; ,
\end{equation}
which defines stationary observers at constant $r$ and $\theta$.
For some region of space-time inside the ergosurface, which is timelike, the vector~(\ref{stationary}) can still be null or timelike. Therefore observers that have a small perturbation in the $r$ direction with respect  to~(\ref{stationary}) can move to increasing $r$.
This implies, like for Kerr, that the ergosurface is not an event horizon. One can then look for the surface inside which stationary observers cease to exist.
In the case of the Kerr metric, this hypersurface is null and it turns out to be the event horizon $r=r_+$.
Moreover, $\omega$ does not depend on $\theta$ at $r=r_+$, and therefore the vector~(\ref{stationary}) is a Killing vector as well, so that the event horizon is also a Killing horizon.

Following the same reasoning as in the case of the Kerr metric, one can write
\begin{equation}
\tilde{g}_{\mu\nu}l^\mu l^\nu =0,
\end{equation}
which results in a quadratic equation in $\omega$ with the solutions
\begin{equation}
\label{omegapm}
\omega_\pm = \frac{1}{\tilde{g}_{\varphi\varphi}} \left(-\tilde{g}_{t\varphi}\pm\sqrt{\tilde{g}_{t\varphi}^2-\tilde{g}_{tt}\tilde{g}_{\varphi\varphi}}\right).
\end{equation}
This second order algebraic equation is fully analogous to its Kerr counterpart.
The solution for~$\omega$ is real only when the discriminant in~(\ref{omegapm}) is positive. This discriminant is nothing but the two dimensional determinant $\text{Det}_{(t,\varphi)}$ of the $(t,\varphi)$ sections of (\ref{df}).
Using~(\ref{df}), we find an equation for the hypersurface where the determinant  $\text{Det}_{(t,\varphi)}=0$, or equivalently the discriminant of (\ref{omegapm}) is zero:
\begin{equation}
\label{P}
P(r,\theta)\equiv \tilde{\Delta}(r) + \frac{2\tilde{M}Da^2r\sin^2\theta}{\rho^2(r,\theta)}=0\, ,
\end{equation}
where $\tilde{\Delta}(r) = a^2+r^2-2\tilde{M}r$.
Note that the same equation can be obtained by requiring that $\tilde{g}^{rr}=0$, which is also similar to the Kerr case, indeed, Eq.~(\ref{rpm}) can also be found from $g^{rr}=0$ of the Kerr metric.
Eq.~(\ref{P}) is a fourth order algebraic equation in $r$, thus one can write an analytic solution for $r$ as a function of $\theta$,
\begin{equation}
\label{Pzero}
P(R_0(\theta),\theta)=0\;\;\; \Rightarrow \;\;\; r=R_0(\theta),
\end{equation}
but we do not give it here since it is not especially informative.
The above equation can have multiple roots, and we are always interested in the outermost one. This resembles the situation with the Kerr metric, where there are at most two solutions.

Let us comment on the hypersurface given by Eq.~(\ref{Pzero}).
First of all, by taking $D=0$, Eq.~(\ref{P}) reduces to the equation determining the locus of the horizon in the Kerr case, $\Delta =0$.
For $D\neq 0$, however, the solution $r=R_0(\theta)$ depends on $\theta$.
This implies that the vector~$l^\mu$ defined in~(\ref{stationary}), although a combination of Killing vectors, is not a Killing vector itself.
Indeed, Eq.~(\ref{omegapm}) taken at the surface $P=0$ yields $\omega(\theta) = -\tilde{g}_{t\varphi}/\tilde{g}_{\varphi\varphi}$, which depends on $\theta$ for $D\neq0$, unlike in the case of Kerr.
Note also that for arbitrary $D$, the surface $P=0$ coincides with the (outer) ergosurface at the poles.
Moreover, in the interior of this hypersurface $P=0$ all Killing vectors are spacelike (except at the poles)!

Another important property of the hypersurface (\ref{Pzero}) is that it is not null.
Indeed, a normal vector to the $P=0$ hypersurface is $N_\mu = \left(0,1, -R'_0(\theta),0\right)$, and its norm is easily evaluated,
\begin{equation}
\tilde{g}^{\mu\nu}N_\mu N_\nu = \tilde{g}^{rr} + R'^2_0\, \tilde{g}^{\theta\theta} >0,
\end{equation}
implying that the hypersurface $r=R_0(\theta)$ is time-like.

The physical meaning of the hypersurface $P=0$ can be seen from its construction: it is the last surface of stationary observers, i.e. observers with constant $\theta$ and $r$. While in the case of the Kerr metric it coincides with the horizon (since it is null), for the disformal metric the hypersurface (\ref{Pzero}) is not a horizon.
By definition, it has the meaning of a stationary limit (c.f. the static limit). Therefore from now on we will refer to the hypersurface $P=0$ as the stationary limit.

\subsection{A candidate event horizon}
The last stationary surface defined by $P=0$ is time-like, as we saw above.
Therefore we have to proceed to even smaller $r$ in order to find a candidate horizon.
The candidate horizon has to be a null hypersurface, is situated inside the stationarity limit and is time independent and axisymmetric according to the metric symmetries~(\ref{df}).
Let us define a normal vector to such a hypersurface $r=R(\theta)$,
\begin{equation}
\label{n}
n_\mu = \left(0,1, -R'(\theta),0\right).
\end{equation}
Asking that the above vector is null, the hypersurface we seek verifies the equation
\begin{equation}
\label{null_surf}
\left(\frac{dR}{d\theta}\right)=\pm \sqrt{-P(R,\theta)}\; ,
\end{equation}
where $P$ is given in~(\ref{P}). A solution to the above equation (\ref{null_surf}), the null hypersurface $R=R(\theta)$, is our candidate event horizon for the disformed Kerr metric.
The above equation can also be found by requiring that the metric is degenerate on $r=R(\theta)$ hypersurfaces, as we demonstrate in  Appendix~\ref{app:geometry}. This is done by defining a new radial variable $\zeta$ such that the horizon is situated at some constant $\zeta$. The same horizon equation (\ref{null_surf}) is then given by $g^{\zeta\zeta}=0$ in a similar way to Kerr in Boyer-Lindquist coordinates. One can check that the vector~(\ref{n}) satisfying (\ref{null_surf}) is a generator of null geodesics skimming the hypersurface $R=R(\theta)$. Eq.~(\ref{null_surf}) with~(\ref{n}) also agrees with the equation defining the horizon in~\cite{Johannsen:2013rqa,Johannsen:2015pca}, albeit for other Kerr deformations.

The differential equation~(\ref{null_surf}) is of first order and has two branches. In each branch, the solution $R=R(\theta)$ is by definition monotonous.
Depending on the sign of $D$ and the interval of $\theta$, one has to choose a particular branch among the two in~(\ref{null_surf}).
Note that Eq.~(\ref{null_surf}) is invariant under the change $R(\pi-\theta)=R(\theta)$, which agrees with the fact that our solution should be symmetric with respect to the equator. For the solution $r=R(\theta)$ to be smooth, its derivatives should vanish at the poles, $R'(0)=R'(\pi)=0$. In addition, since we ask for the solution to be symmetric with respect to $\theta\to \pi-\theta$, we have to also require that $R'(\pi/2)=0$, which is not automatic. Note that by virtue of Eq.~(\ref{null_surf}) these conditions mean that the surfaces $R(\theta)$ and $R_0(\theta)$ coincide at the poles and the equator.

\begin{figure}[t]
    \centering
    \includegraphics[scale=0.8]{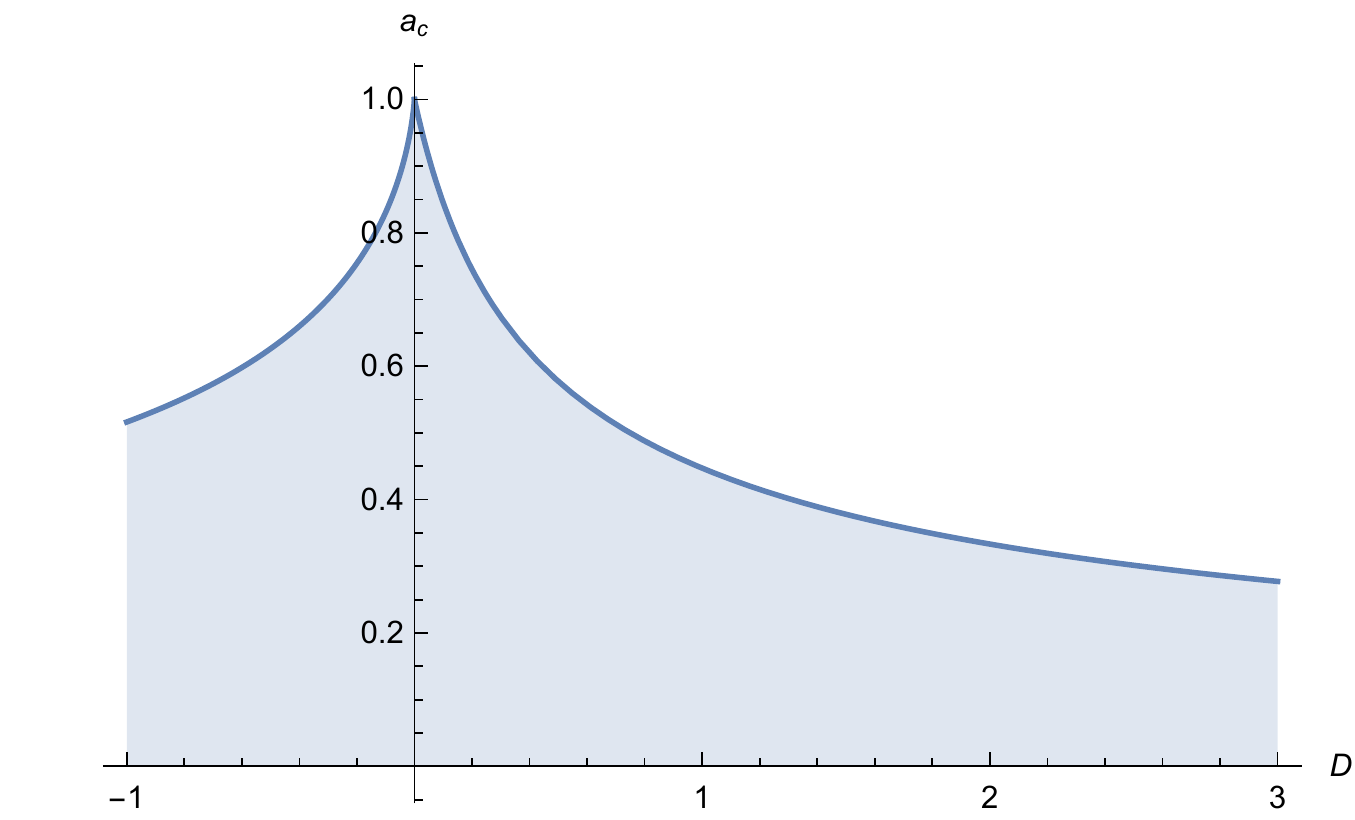}
    \caption{Critical value $a_c$ as a function of $D$. For $a>a_c$, we have $R'(\pi/2)\neq0$ when $D<0$, and $R'(0)\neq0$ when $D>0$. For points in the shaded region, $R'(0)=R'(\pi/2)=0$ is allowed.}
    \label{a_crit}
\end{figure}

On the other hand, the equation~(\ref{null_surf}) is of first order, therefore we only need to specify one boundary condition.
This implies that {\it a priori} there are more physical requirements on the solution than the available freedom in the choice of the boundary conditions. As we will see by solving~(\ref{null_surf}) numerically, this indeed becomes a problem for some ranges of the parameters $\tilde{M}$, $a$ and $D$.
It is possible to see analytically why one of the conditions on the solution of (\ref{null_surf}) cannot be satisfied for large enough values of $a$ and $|D|$.
To do this, we examine the behaviour of the solution around $\theta=0$ and $\theta=\pi/2$.  For the sake of convenience, we discuss the results in terms of `natural' units setting $\tilde{M}=1$. This means that the radius $r$ (as well as $R_0(\theta)$ and $R(\theta)$) and the Kerr parameter $a$ are given in terms of $\tilde{M}$. We assume that $R$ is twice differentiable at $\theta=\pi/2$. Then, requiring that $R'(\pi/2)=0$,
a Taylor expansion around $\theta=\pi/2$ yields a necessary condition,
\begin{equation}
\label{real_cond}
\left(R-1 - \frac{Da^2}{R^2}\right)^2\geq -8 Da^2\frac{R^2+a^2}{R^3}\; ,
\end{equation}
in order for $R''(\pi/2)$ to be real,
where in the above equation $R$ is evaluated at $\theta=\pi/2$.
Eq.~(\ref{real_cond}) is automatically satisfied for positive $D$, but not  for negative $D$.  In this case, substituting the relevant solution for $R=R_0$ at $\theta=\pi/2$ into (\ref{real_cond}),  one can show that $a$ cannnot exceed a critical value $a_c(D)$ which verifies a fourth order polynomial equation in $a_c^2$,
\begin{equation}
\label{acrit_negative}
Q_3 (a_c)=0\; ,\qquad D<0\; ,
\end{equation}
where the expression for $Q_3$ is given in Appendix~\ref{AppendixQ}.
The above equation can be solved in terms of $D$ (one can show that there is only one positive real solution).
For $D>0$, similar arguments at $\theta=0$ yield
\begin{equation}
\label{acrit_positive}
a_c = \frac{1}{\sqrt{1+4D}}\; ,\qquad D>0\; .
\end{equation}
The solution for $a_c$ as a function of $D$ is shown in Fig.~\ref{a_crit}. In terms of the angular momentum $\tilde{a}$ measured by an observer at infinity (see section \ref{sec:properties}), one has $\tilde{a}<\sqrt{1+D}\,a_c$ for physical solutions. Using \ref{acrit_negative} and \ref{acrit_positive}, one can check that $\tilde{ a}<1$ when $D\neq0$, meaning that the disformed metric looks like a sub-extremal Kerr solution to an observer at infinity.

\begin{figure}[h]
    \centering
    \includegraphics[scale=1.1]{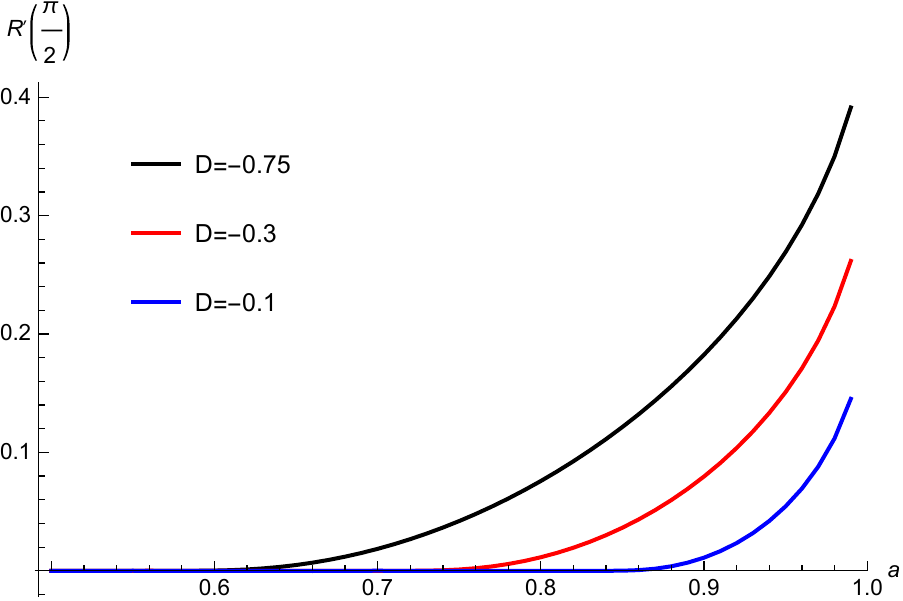}
    \caption{Numerical value for $R'(\pi/2)$ for $D=-0.75$ (black), $D=-0.3$ (red) and $D=-0.1$ (blue).}
    \label{derivative_equator}
\end{figure}

Note that this study allows us to verify that values of $a> a_c(D)$ do not give a smooth solution for the null surface\footnote{Higher orders of the Taylor expansion around $\theta=\pi/2$ do not give additional conditions. Indeed, if we assume that $R^{(2p+1)}(\pi/2)=0$, then the order $2p$ in the Taylor expansion is linear in $R^{(2p)}$ when $p>1$. So we do not have additional constraints to ensure that $R^{(4)},R^{(6)}$, etc. are real at $\theta=\pi/2$. The same is true for the expansion around $\theta=0$.}.
This does not guarantee, however, that the solution is smooth for  $a<a_c$. We could only verify this numerically, as can be seen from Fig.~\ref{derivative_equator}. The numerical integration yields $R'(\pi/2)=0$ to a high precision when $a<a_c$, and $R'(\pi/2)\neq0$ when $a>a_c$. Similar results are obtained when evaluating $R'(0)$ for $D>0$.

As for finding the solution of~(\ref{null_surf}), we used two different techniques.
The direct numerical integration of~(\ref{null_surf}) with a given boundary condition by the Runge–Kutta method is discussed below.
An alternative approach is to search for the solution $r=R(\theta)$ by expanding~(\ref{null_surf}) in an iterative series,
where at each order a solution can be found knowing the result at the previous order.
We refer the reader to Appendix~\ref{sec:small_D} for this approach.

Depending on the sign of $D$, it is convenient to choose the boundary condition either at $\theta=0$ or $\theta=\pi/2$ in our numerical integration.
Having this in mind, we only have to solve Eq.~(\ref{null_surf}) in either one of the two intervals $0\leq\theta\leq\pi/2$ or $\pi/2\leq\theta\leq\pi$. We are again using natural units ($\tilde{ M}=1$) for the numerical results.

\begin{figure}[t]
\centering
\includegraphics[scale=1.]{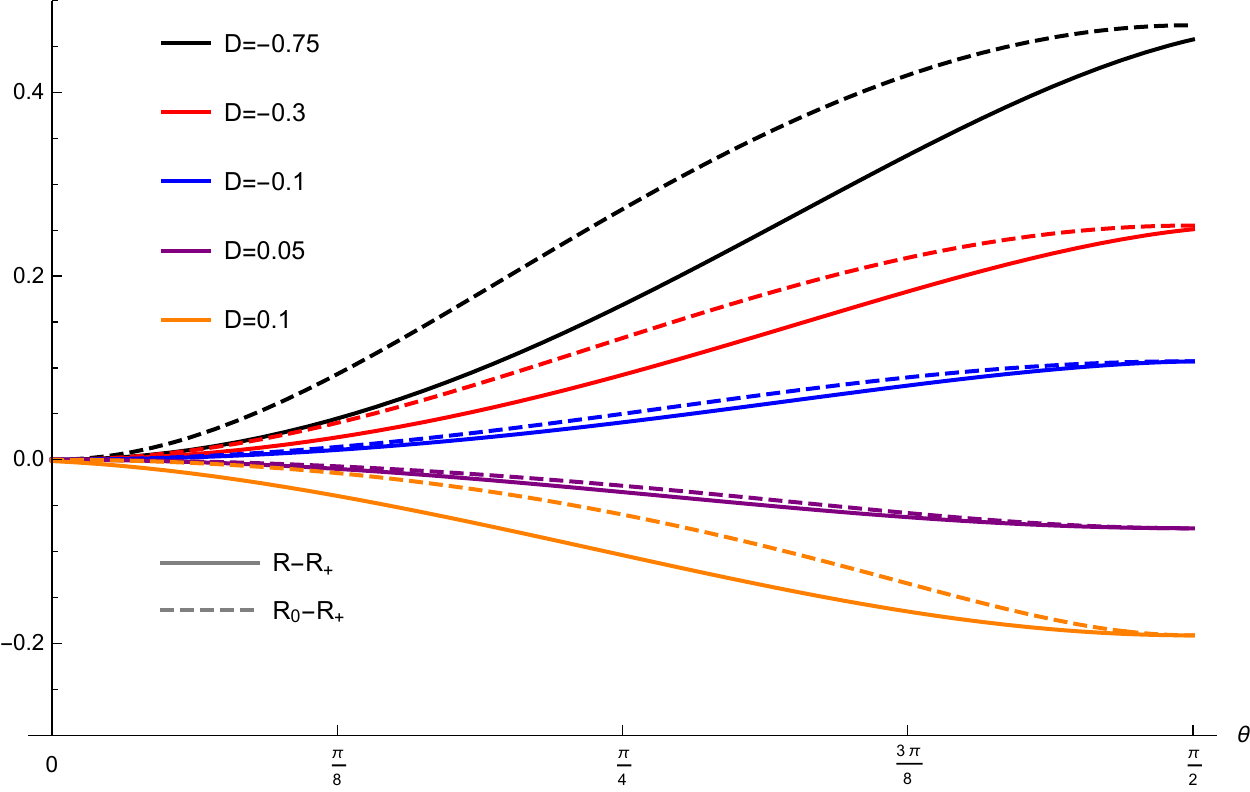}
\caption{Numerical integration of $R-R_+$ and $R_0-R_+$ for $a=0.9$ and varying $D$, respectively $D=-0.75$ (black), $D=-0.3$ (red), $D=-0.1$ (blue), $D=0.05$ (purple) and $D=0.1$ (orange). The solution becomes unphysical when $|D|$ becomes large.}
\label{DX02}
\end{figure}

Let us first consider the case $D<0$.
The physical branch of (\ref{null_surf}) is increasing for $0\leq\theta\leq\pi/2$ and decreasing for $\pi/2\leq\theta\leq\pi$, so that the solution $r=R(\theta)$ reaches its maximum at the equator.
In this case we numerically integrate the equation for $R'(\theta)$ in the interval $0\leq\theta\leq\pi/2$\footnote{Note that our numerical integration fails when we try to integrate in the other range, for $\pi/2\leq\theta\leq\pi$. We believe that this is due to the numerical instability exploding when the negative branch is chosen in~(\ref{null_surf}).}.
For small values of $a$, the curves $R(\theta)$ and $R_0(\theta)$ are extremely close in the whole range $[0,\pi/2]$, and the numerical results are consistent with the condition $R'(\pi/2)=0$.
This remains true as we increase the rotation, until the rotation parameter reaches a critical value $a=a_c$.
For higher values $a>a_c$, we have $R'(\pi/2)\neq0$, which is clear from the fact that $R\neq R_0$ at the equator (see Fig.~\ref{DX02}).

In the case $D>0$, the physical branch of the solution $r=R(\theta)$ is decreasing for $0\leq\theta\leq\pi/2$ and increasing for $\pi/2\leq\theta\leq\pi$. Thus the null surface has a minimum at the equator, contrary  to the case $D<0$.
Apart from the constraint on $a$ coming from the consistency of~(\ref{null_surf}) at $\theta=\pi/2$ discussed above, in this case there is an additional bound on the possible values of $a$ and $D$.
Indeed, for both  $a$ and $D$ large, there is no longer a solution of $P(r,\pi/2)=0$, which means that the condition $R'(\pi/2)=0$ cannot be satisfied.
In other words, for a given $D$, an increase of $a$ results in the disappearance of the root of $P$ at the equator.
However, one can show that  the bound coming from the existence of $R_0(\pi/2)$ is weaker than the one obtained in (\ref{acrit_positive}).
On the other hand, if both $D$ and $a$ are not too large, the numerical solution of~(\ref{null_surf}) is smooth at $\theta=\pi/2$, see Fig.~\ref{DX02}.
Note that in this case we solve from $\theta=\pi/2$ to $\theta=\pi$, since the numerical integration breaks down in the interval from $0$ to $\pi/2$. Similarly to the case of negative $D$ in the interval $[\pi/2,\pi]$, we believe that this is related to a growth of the numerical  instability.

Having found the null hypersurface $R(\theta)$, let us comment on the causal structure of spacetime. 
We consider the following continuous one parameter family of hypersurfaces, $R_\zeta(\theta)=R(\theta) +\zeta$, where $\zeta$ is a constant labeling a given hypersurface. 
The parameter $\zeta$ corresponds to the coordinate $\zeta$ introduced in the Appendix~\ref{app:geometry}, i.e. the surfaces $R_\zeta(\theta)$  are $\zeta=\text{const}$ surfaces in coordinates $(t,\zeta,\theta,\varphi)$. These coordinates are adapted to the candidate horizon situated at $\zeta=0$, whose equation is simply $g^{\zeta\zeta}=0$, which is identical to (\ref{null_surf}).
In other words the set of hypersurfaces $R_\zeta(\theta)$ is analogous to $r=\text{const}$ surfaces in the Kerr case. 
Therefore $\zeta=0$ corresponds to the null surface $R(\theta)$, while for $\zeta>0$ ($\zeta<0$) the surface $R_\zeta(\theta)$ is outside (inside) the candidate horizon. Under the mild assumption that $R(\theta)> \tilde{M}$ (non-extremality for Kerr), it can be shown that these surfaces are timelike for $\zeta>0$, and that there exists small $\zeta_0<0$, such that the $R_\zeta(\theta)$ are spacelike for $\zeta_0<\zeta<0$. 
This construction shows that outside of $R(\theta)$ null (and timelike) geodesics can travel towards increasing {\it and} decreasing $\zeta$, crossing the hypersurfaces $R_\zeta(\theta)$ both ways. On the other hand, inside the candidate horizon surface, null (and timelike) geodesics can only move in one direction. This picture is fully analogous to the Kerr black hole for radii greater than the inner horizon radius. We therefore conclude that under the assumption that a regular solution $R=R(\theta)$ exists, our candidate horizon is indeed an event horizon shielding the ring singularity. The disformed Kerr metric is therefore a black hole.

It may seem paradoxical that inside the surface of last stationary observers, there is still a region of space-time from which light and particles can escape. Indeed, the stationarity vector (\ref{stationary}) (which we defined with constant $\theta$ and $r$ coordinates) is spacelike inside the stationary limit.
However, timelike vectors with {\it non-constant} $\theta$ and $r$ can be constructed in between the surfaces $r=R_0(\theta)$ and $r=R(\theta)$. Intuitively, this can be understood from the fact that there is a non-zero $\tilde{g}_{tr}$ term compared to the Kerr metric in Boyer-Lindquist coordinates. This additional term plays a complementary role to the $g_{t\phi}$ term in Kerr, which distinguishes static from stationary observers. Here, it allows the existence of timelike observers moving towards increasing $\zeta$ within the stationarity limit.

\section{Discussion, Conclusions}

In this article we have considered a disformal transformation of the
standard GR Kerr metric. Crucially, the disformal directions were
given with respect to derivatives of the scalar field which are
tangent vectors to a regular geodesic congruence of the spacetime
metric. The scalar is a particular Hamilton Jacobi function for Kerr
geodesics. The four conserved parameters of the Kerr spacetime
(originating from the two Killing vectors and the two Killing
tensors{\footnote{Here, we include the metric itself which is trivially a Killing tensor for a metric connection.}}) are chosen so that the scalar is well defined from the
event horizon to asymptotic infinity \cite{Charmousis:2019vnf}. The
resulting disformal metric is a stationary and axisymmetric spacetime, like Kerr, while the
scalar field is again related to the disformed spacetime geodesics.
We have shown that the resulting spacetime is non circular and non
Einstein (unlike Kerr), but that it has a single ring singularity
at $\rho=0$, just like Kerr. 
We have found compelling
evidence (an important number of necessary conditions) that there
exists a regular null hypersurface, our event horizon,
situated at $R=R(\theta)$ (solution of (\ref{null_surf})). This
hypersurface was shown to be situated in the interior of the
stationary limit of constant $r$ and $\theta$ observers. This latter stationary limit surface is given by an equation $P(r,\theta)=0$,
(\ref{P}), and is located inside the ergosurface, which is given by $\tilde{ g}_{tt}=0$.
We have shown that for large $r$, the disformal metric resembles the Kerr solution with a rescaled mass $\tilde{ M}$ and angular momentum $\tilde{a}$. Furthermore, asking for the event horizon to be physical results in an upper bound for the rotation parameter $\tilde{a}$, and one can show that $\tilde{a}<\tilde{ M}$ if $D\neq0$, meaning that the disformed solution will look like a sub-extremal Kerr spacetime to an observer at infinity.

In summary, we have found necessary conditions and numerical evidence for the existence of a regular null hypersurface $R=R(\theta)$ which is regular at the equator (no knee singularity). We showed that when this hypersurface is present it is the boundary of a trapped interior region from where no lightlike (or timelike) signals can escape (in a stationary spacetime). Under the assumption of a regular $R=R(\theta)$ surface, we have therefore shown that our candidate horizon is indeed the event horizon of a black hole while demonstrating that our spacetime is free of time machines (causally stable). The latter is a rather nice feature  due to the presence of a global time function $\phi(t,r)$, inherent to the construction of the disformal metric~(\ref{df}) using timelike geodesics.

The disformal transformation is an internal map within DHOST Ia theories. We start from $c_T=1$ theories where
our spacetime is identical to the GR Kerr solution \cite{Charmousis:2019vnf}, and map to a disformed Kerr metric for
some DHOST Ia theory with some given $G$, $A_3$ and $A_1$ parameters in the notation of DHOST \cite{Achour:2016rkg}. Such theories
are constrained from gravity wave tests (see for example \cite{Ezquiaga:2018btd}) assuming that the scalar is varying at vast
cosmological scales ie., a dark energy field.{\footnote{There have been criticisms on such effective theory calculations that
recent data from LIGO/Virgo are within the strong coupling scale \cite{deRham:2018red} associated to dark energy and one has
to be rather careful when making stringent claims.}} The solutions we have discussed here are asymptotically flat and locally influence the speed of gravity waves for these particular scalar tensor theories. Independently of gravity wave constraints, the solutions
discussed here go beyond the interest of these particular theories and we believe that they are interesting in their own right as simple,
analytic, benchmark alternatives to the prototype Kerr solution. 
Indeed, one may consider the $D$ dependent metric~(\ref{df}) as a one-parameter  family of Kerr deformations, which may be tested by  present and future gravity experiments. 
In the metric~(\ref{df}), deviations from GR are encoded in the disformality coefficient $D$, therefore one may look either for constraining deviations from GR or, on the other hand, for smoking gun gravity modifications.
Furthermore, given their simple origin related to geodesics,
they may include effects beyond probable strong coupling scales of particular EFT theories.  In this sense we think it would even
be very interesting to study disformations of dark energy self tuning solutions starting from the regular solutions
in \cite{Charmousis:2019vnf}. Additionally, one could include conformal transformations which will not alter the null cones
but may yield interesting regularity conditions \cite{Domenech:2019syf}.

The presence of the non circular off-diagonal
term $\tilde{g}_{tr}$ can  also be instructive when looking for approximate
solutions, for example in the slowly rotating limit. Indeed, for the disformal solution the resulting off-diagonal term can be eliminated by a
change of coordinates in the first order approximation, but this is
no longer the case from the next order on. This observation may be
pertinent for the search of a rotating black hole solution of the
so-called Chern-Simons modified gravity \cite{Jackiw:2003pm}. For
example, in Ref.\cite{Grumiller:2007rv}, the authors claim that
stationary and axisymmetric solutions of Chern-Simons modified
gravity probably do not exist, but they limit their analysis to
circular configurations. In order to bring light to this issue, one must clearly include metric contributions that are not circular. Signs of non-circularity 
have also been found numerically in the case of a DGP Horndeski rotating black hole \cite{VanAelst:2019kku}. 
Breaking the circularity hypothesis constitutes a milder approach to the one concluded in \cite{Grumiller:2007rv},
where it is claimed that the related spinning black hole should break either the
stationarity or axisymmetry hypothesis (or both). Our relatively simple analysis hints that the circularity hypothesis is very much tied in with Einstein metrics and GR, but most probably not modified gravity. 
Therefore circularity should not be taken for granted for axially symmetric and stationary metrics beyond the realm of GR.

Last but not least, the solutions described here are interesting on
purely theoretical grounds as counterexamples to usual GR black hole
metrics.  
Indeed, there are numerous questions which we have left unanswered, starting with the global causal structure of the spacetime~(\ref{df}). This question might be rather non-trivial and requires a separate study.
Furthermore, there are issues related to the failure of the event horizon to be a Killing horizon. 
Classical black hole theorems which
assert that an event horizon is also a Killing horizon for an
axially symmetric and stationary spacetime~\cite{Carter:1971zc} fail due to the non circularity of spacetime.
Therefore, how can one define surface gravity here and how can one go
about studying the thermodynamics of these solutions ? The possibility of extending the notion of surface gravity for horizons which are no longer Killing has been
studied by several authors in different contexts (see \cite{Cropp:2013zxi} and references within). It would be interesting to study the thermodynamics of this specific solution under the differing definitions provided for surface gravity.
Other interesting questions include: do we still have an additional Killing tensor for this
disformed metric and are geodesics integrable, or  what is the
effect of having an extra ergosurface on the Penrose process ? These are some of the interesting questions that one
can consider starting from this relatively simple construction.

{\it Note added.} A study on a similar subject appeared recently~\cite{BenAchour:2020fgy}, where, amongst other findings,  the authors confirmed
some of the results presented here.

\section*{Acknowledgements}
We are very happy to thank Eloy Ay\'on-Beato, Victor Berezin, Marco Crisostomi, Vyacheslav Dokuchaev, Yury Eroshenko, Eric Gourgoulhon, Karim Noui, George Pappas, Scott Robertson and Alexey Smirnov for interesting discussions.
The work was supported by the CNRS/RFBR Cooperation program for 2018-2020 n. 1985 ``Modified gravity and black holes: consistent models and experimental signatures''.
CC acknowledges support from
the CNRS grant 80PRIME and warmly thanks the Laboratory
of Astronomy of AUTh in Thessaloniki for hospitality
during the course of this work and in particular the virus outbreak. The authors also gratefully acknowledge the kind support of the ECOSud project C18U04.


\appendix

\numberwithin{equation}{section}

\section{The disformed metric and the scalar field in regular coordinates}
\label{AppendixKerr}
The disformed Kerr metric~(\ref{df}) inherits the problem from the Boyer-Lindquist presentation of the Kerr solution, namely that it has a singularity at $r=r_\pm$.
While it is well known that in the case of the GR solution~(\ref{BL}), this singularity is not physical, we {\it a priori} do not know whether this is also a coordinate singularity in the disformed metric~(\ref{df}).
Moreover, since the scalar field is a part of the modified gravity theory, we would also like to establish that the scalar is regular at $r=r_\pm$, although Eq.~(\ref{phi}) may suggest otherwise.
To see explicitly that both the metric and the scalar are regular at $r=r_\pm$, we repeat the calculations of Sec~\ref{sec:disformal}, starting from a regular form of the Kerr metric. The metric of a rotating black hole in GR can be written in the Kerr coordinates~\cite{Kerr:1963ud}:
\begin{equation}
\label{Kerr}
\td s^2 = -\left(1-\frac{2Mr}{\rho^2}\right)\left(\td v + a\sin^2\theta \td \vp\right)^2 + 2\left(\td v + a\sin^2\theta \td \vp\right)\left(\td r + a\sin^2\theta \td \vp\right) + \rho^2\left(\td\theta^2 + \sin^2\theta \td \vp^2\right)\; ,
\end{equation}
which, unlike the same metric in Boyer-Lindquist coordinates, does not have a singularity at $r=r_\pm$.
The connection to the Boyer-Lindquist coordinates is made via
\begin{equation}
\begin{split}
t &\to v - r - \int\frac{2M r}{\Delta}\td r\;, \\
\vp & \to -\vp -a\int \frac{\td r}{\Delta}\; ,
\end{split}
\end{equation}
In the Kerr  coordinates, the scalar field reads
\begin{equation}
\label{phiKerr}
\phi = q\left(v-r+\int\frac{\td r}{1+\sqrt{\frac{r^2+a^2}{2Mr}}}\right)\; ,
\end{equation}
and one sees that it is regular.
Applying the disformal transformation~(\ref{disformal}) to the metric~(\ref{Kerr}) and using the expression for scalar field~(\ref{phiKerr}), one can straightforwardly obtain the disformal metric in the Kerr-like coordinates,
\begin{equation}
\begin{split}
\label{dfKerr}
\td\tilde{ s}^2 &= -\left(1+D -\frac{2Mr}{\rho^2}\right) \td v^2 + 2\left( 1 + D -\frac{D}{1+\sqrt{\frac{r^2+a^2}{2Mr}}}\right)\td v\td r -D \left(1-\frac{1}{1+\sqrt{\frac{a^2+r^2}{2Mr}}}\right)^2 \td r ^2\\
& + \frac{4 a M r \sin^2\theta}{\rho^2}\td v \td \vp + 2 a \sin^2\theta \td r \td \vp + \rho^2 \td \theta^2 \\
&+ \frac{\sin^2\theta\left(2 a^4\cos^2\theta + 4 a^2 M r \sin^2\theta+ a^2 r^2\left[3 + 2 \cos(2\theta)\right] + 2 r^4\right)}{2 \rho^2}\td \vp^2\; .
\end{split}
\end{equation}

\section{Polynomials $Q_i$}
\label{AppendixQ}
For completeness, we list here the polynomials that we encountered in the main text.
In Sec.~\ref{sec:properties} the two polynomials $Q_1(r,\theta)$ and $Q_2(r,\theta)$ appear in the expressions for the curvature invariants:
\begin{align*}
Q_1(r,\theta)&= \left[127 + 56 \cos(2\theta) + 9 \cos(4\theta)\right]r^4+  4 a^2\left[33+ 14 \cos(2\theta) + \cos(4\theta)\right]r^2 +18 a^4 \sin^2(2\theta)  \; ,\\
Q_2(r,\theta)&= 48\left(r^2+a^2)\left(r^6-15a^2 r^4 \cos^2\theta  + 15a^4 r^2\cos^4\theta - a^6 \cos^6\theta\right)\right) \\
& \,-\frac{D a^2}{2} \large\{160\left[4 + 3 \cos(2\theta)\right]r^6 - a^2\left[3\left(52+3D\right)\cos(4\theta) + 4\left(48+25D\right)\cos(2\theta)-124+243 D\right]r^4\\
&\, + a^4 \left[3 \cos(6\theta) + \left(D-138\right)\cos(4\theta)-\left(627+100D\right)\cos(2\theta)-486-253D\right]r^2\\
&\, + 12 a^6 \cos^2\theta\left[\cos(4\theta) + \left(4 + 6 D\right)\cos(2\theta) + 3-6D\right]\large\}\; .
\end{align*}
In Sec.~\ref{sec:surfaces} the polynomial equation determining the critical value of $a$ for negative $D$ appears with the following expression for $Q_3(a_c)$:
\begin{gather*}
Q_3(a_c)=-256 D^2 + 32 D\left[39 + D\left(50-13D\right)\right]a_c^2 + \left[15 + D\left(343 D^3 + 2324 D^2 + 562 D -2076\right)\right]a_c^4\\ - 2 \left[15- D\left(414 + 517 D\right)\right]a_c^6 + 15 a_c^8.
\end{gather*}

\section{Geometry of the null surface}

\label{app:geometry}

When $D=0$, Eq.~(\ref{null_surf}) is compatible with constant $R$
solutions, which respectively read $R_\pm =
\tilde{M}\pm\sqrt{\tilde{M}^2-a^2}$. These correspond to the
horizons of the Kerr solution with a rescaled mass. A necessary
condition to have a solution to \ref{null_surf} is $P(R,\theta)\leq
0$. We set $R'(0)=0$ in order for the solution to be regular at the
north pole. This implies that $P(R(0),0)=0$, which has two solutions
$R(0)=R_\pm$. We choose $R(0)=R_+$ in order to have a common locus
with the ergosphere.  We can focus on half of the interval,
$\theta\in[0,\pi/2]$, since the differential equation defining the
horizon is symmetric under $\theta\to\pi-\theta$. In fact, $R'$ changes sign in the
second half interval as we change branch in the
differential equation. In order to have a regular solution, one must
also have $R'(\pi/2)=0$. Indeed, at $\theta=\pi/2$, we need
$R'(\pi/2)=0$ so that our junction at the equator has a continuous
first derivative to the second branch in a regular fashion for
$[\pi/2, \pi]$. In between $\theta=0$ and and $\theta=\pi/2$ we have
that $R(\theta)$ is non trivial, monotonous  and therefore we must
have that $\tilde{\Delta}+\frac{2\tilde{M}D R a^2
\sin^2\theta}{\rho^2}$ is strictly negative. Note that we cannot
impose $R'(\pi/2)=0$, since we have already used up our initial
condition upon setting $R'(0)=0$. The condition $R'(\pi/2)=0$ is
however compatible with the differential equation, as one can show by
calculating the derivative of (\ref{null_surf}) at $\theta=\pi/2$,
\begin{equation}
\left(\frac{dR}{d\theta}\right)\left(\frac{d^2R}{d\theta^2}+R-\tilde M -\frac{\tilde M a^2 D}{R^2}\right)=0,\qquad \mbox{at} \qquad \theta=\pi/2
\end{equation}
but not unique. If the solution singles out the second branch pictured above, there will be a knee singularity at the horizon's equator. As we will see this depends on the magnitude of $a$ and the magnitude and sign of $D$.\footnote{A similar branching occurs at $\theta=0$ but there we can choose the $R'=0$ branch via our unique initial condition. The only possible zeros for $R'$ occur at $\theta=0$ and $\theta=\pi/2$ in the half interval $[0,\pi/2]$.}

Unlike for the Kerr metric, the horizon $r=R(\theta)$ is not at constant $r$ coordinate.
This is of course coordinate related and in the following we will define a radial coordinate $\zeta$ which is adapted to the horizon.
 We can set $H=\sqrt{-(\tilde{\Delta}+\frac{2\tilde{M}D R a^2 \sin^2\theta}{\rho^2})}$. In other words we have $H\leq 0$ at the horizon surface. We can now define $d\zeta=dr-\epsilon H(\theta) d\theta$ as a new radial coordinate which is such that the horizon is now at $d\zeta=0$, i.e. at some constant radial coordinate $\zeta=\zeta_0$, and $\epsilon=\pm 1$ labels the branch fixing the sign of $R'$. The coordinate transformed metric replacing $r$ by $\zeta$ has even more cross-terms and is not particularly helpful given that we do not explicitly know the function $R$. We can however set $d\zeta=0$ and write down the 3-dimensional hypersurface representing the event horizon geometry. It is easy to verify that in $\zeta$ (instead of $r$) coordinates our horizon equation (\ref{null_surf}) is accordingly given by $g^{\zeta\zeta}=0$.
 This hypersurface  is obtained by setting $r=R(\theta)$ and $dr=\epsilon H(\theta) d\theta$. By direct substitution and after some calculation, the metric reduces to a perfect square:
\begin{gather}
\label{disformal_metric31}
ds^2_3 = \frac{\left[\td \Theta^2+\sin^2\theta (2aMR \td t-{\cal{B}}\td \varphi)^2 \right]}{(1+D){\cal{B}}\rho^2}\; ,
\end{gather}
where $$\td \Theta=\sqrt{\frac{2MR(a^2+R^2) D^2}{\Delta^2}}{\cal{B}} \td \theta-\epsilon \sqrt{[2MR(R^2+a^2)-(1+D){\cal{B}}]\rho^2} \td t$$ and ${\cal{B}}=(R^2+a^2)\rho^2+2MRa^2 \sin^2\theta$, and the angle $\vp$ has been rescaled as $\vp\to\vp/\sqrt{1+D}$. We can note that the $\td t^2$ coefficient is positive, which is to be expected since we are within the ergoregion and $\partial_t$ is spacelike. The determinant of the metric is found to be zero and we have a null hypersurface. The (perfect square) form of the metric suggests that for the null vector generators we take $L^a=\{1,L^\theta,L^\phi\}$ such that $g_{\mu\nu}^3 L^\mu L^\nu=0$.
Putting it all together we get:
\begin{gather}
\label{generator}
L^\theta=\sqrt{\frac{-\Delta^2 \rho^2 (\rho^2 \Delta +D {\cal{B}})}{2MR D^2 (R^2+a^2) {\cal{B}}^2}}\; , \qquad L^\varphi=\frac{2MR a}{{\cal{B}}}\; ,
\end{gather}
where we set $L^t=1$ without loss of generality and $R=R(\theta)$ is a solution of the differential equation (\ref{null_surf}).
There is a set of null curves, defined on the hypersurface $g^{\zeta\zeta}=0$, which are tangent to the four dimensional trivial extension of $L^a=\{1,0,L^\theta,L^\phi\}$ which is a null vector defined throughout this hypersurface. These 4-dimensional null curves are defined via $\frac{dY^a}{d\lambda}=L^a$ with $Y^a=\{\lambda,\zeta_0, \lambda L^\theta+\theta_0, \lambda L_\phi+\phi_0\}$  where $\lambda$ is some parameter for the curve.  These curves describe the last photon orbits skimming the null hypersurface without falling in nor falling out. This is the candidate event horizon for our disformal black hole. The vector $L^a$ is  not a Killing vector, however it coincides with the Kerr Killing generator $L_{Kerr}^a=\{1,0,\frac{a}{R_+^2+a^2}\}$ for $D=0$ and also at the poles for $D\neq 0$. If we set $R'(\pi/2)=0$ at the equator, as we need a smooth solution,  we have $2MR_\star=\frac{(1+D)R_\star^2(R_\star^2+a^2)}{R_\star^2-D a^2}$, where $R(\pi/2)=R_\star$. Then we see that, $R_\star^2 \Delta +D {\cal{B}}=0$ and hence $L^\theta=0$ at the equator, while $L^\phi=\frac{(1+D)a}{R_\star^2+a^2}$. This is consistent with the Kerr case, for which we have $D=0$ and $R_\star=R_+$ (see for example \cite{Visser:2007fj} for a similar discussion on the Kerr metric). Then, switching to the second branch, we can move to the second half interval and the three dimensional metric is again smooth and well defined until we reach the south pole at $\theta=\pi$, where our generator again becomes Killing.

\section{Solving the equation for the candidate horizon via a perturbation expansion}

\label{sec:small_D}

In Sec.~\ref{sec:surfaces}, we discussed the numerical integration of equation~(\ref{null_surf}) for the null hypersurface.
In this Appendix, we attack the same problem with a different approach. Using
(\ref{P}), Eq.~(\ref{null_surf}) can be rewritten as follows:
\begin{equation}
\label{null_surf2}
\left(\frac{dR}{d\theta}\right)^2+ R^2+a^2-2R + \frac{2Da^2 R \sin^2\theta}{R^2+a^2\cos^2\theta} =0\; ,
\end{equation}
where we use  natural units $\tilde{M}=1$.
For small $Da^2$ the last term on the left-hand side can be considered as a perturbation.
We can then write the solution of~(\ref{null_surf2}) as a perturbative series with the perturbation parameter $Da^2$:
\begin{equation}
    \label{expansion}
    R = R_+ + \sum_{n=1}^\infty (Da^2)^n \delta R_n,
\end{equation}
where $R_+=1 + \sqrt{1-a^2}$ is the outer horizon for the Kerr metric with rescaled mass $\tilde{M}$.
Substituting~(\ref{expansion}) in (\ref{null_surf2}), we obtain an algebraic equation for $\delta R_n(\theta)$ at each order, expressed through $\delta R_n(\theta)$  and $\delta R'_n(\theta)$.
In particular, the first correction in the expansion reads
\begin{equation}
    \delta R_1(\theta) = -\frac{\sin^2\theta \left(1 + \sqrt{1-a^2}\right)}{2(1-a^2)+(2-a^2\sin^2\theta)\sqrt{1-a^2}}\; .
\end{equation}
The deviations from the Kerr solution are maximal at the equator, and are of order $R(\pi/2)\sim |Da^2|$ (for small $Da^2$).
Notice that the sign of the leading correction $\delta R_1$ is opposite to the sign of $D$, which is in accordance with our findings in Sec.~\ref{sec:surfaces}.

Yet another version of the same approach is to make an expansion around the solution of $P=0$, i.e. around $R_0(\theta)$.
This is motivated by our numerical results of Sec.~\ref{sec:surfaces}, where we showed that the solution $r=R(\theta)$ is close to $R_0(\theta)$, see Fig.~\ref{DX02}. Therefore, it is natural to look for a small deviation around $R_0(\theta)$.
Since in this case the would-be small parameter $Da^2$ is already present in the zeroth-order solution $R_0(\theta)$,
one should use another approach.
Assuming that the solution is known  at $n$-th order in the approximation, the solution at the next iteration is,
\begin{equation}
R_{n+1}= R_n - \frac{R'^2+P(R)}{\left(\partial P/\partial R\right)}\bigg|_{R=R_n(\theta)}.
\end{equation}
The last expression is obtained by performing a Taylor expansion of Eq.~(\ref{null_surf2}) around $R_n(\theta)$ and neglecting $(R'_{n+1}-R'_n)$ with respect to $R'_n$. Similarly to the approach considered above, we obtain an algebraic equation at each step.

\providecommand{\href}[2]{#2}\begingroup\raggedright\endgroup

\end{document}